\documentclass[twocolumn,preprintnumbers,amsmath,amssymbm,prd]{revtex4}
\usepackage{epsfig}
\usepackage{graphicx}

\begin{document}
\title{Dragging of inertial frames in the composed black-hole-ring system}
\author{Shahar Hod}
\address{The Ruppin Academic Center, Emeq Hefer 40250, Israel}
\address{ }
\address{The Hadassah Institute, Jerusalem 91010, Israel}
\date{\today}

\begin{abstract}
\ \ \ A well-established phenomenon in general relativity is the
dragging of inertial frames by a spinning object. In particular, due
to the dragging of inertial frames by a ring orbiting a central
black hole, the angular-velocity
$\Omega^{\text{BH-ring}}_{\text{H}}$ of the black-hole horizon in
the composed black-hole-ring system is {\it no} longer related to
the black-hole angular-momentum $J_{\text{H}}$ by the simple
Kerr-like (vacuum) relation
$\Omega^{\text{Kerr}}_{\text{H}}(J_{\text{H}})=J_{\text{H}}/2M^2R_{\text{H}}$
(here $M$ and $R_{\text{H}}$ are the mass and horizon-radius of the
black hole, respectively). Will has performed a perturbative
treatment of the composed black-hole-ring system in the regime of
{\it slowly} rotating black holes and found the explicit relation
$\Omega^{\text{BH-ring}}_{\text{H}}(J_{\text{H}}=0,J_{\text{R}},R)=2J_{\text{R}}/R^3$
for the angular-velocity of a central black hole with zero
angular-momentum, where $J_{\text{R}}$ and $R$ are respectively the
angular-momentum of the orbiting ring and its proper circumferential
radius.
%\newline
Analyzing a sequence of black-hole-ring configurations with
adiabatically varying (decreasing) circumferential radii, we show
that the expression found by Will for
$\Omega^{\text{BH-ring}}_{\text{H}}(J_{\text{H}}=0,J_{\text{R}},R)$
implies a {\it smooth} transition of the central black-hole
angular-velocity from its asymptotic near-horizon value
$\Omega^{\text{BH-ring}}_{\text{H}}(J_{\text{H}}=0,J_{\text{R}},R\to
R^{+}_{\text{H}})\to 2J_{\text{R}}/R^3_{\text{H}}$ (that is, just
{\it before} the assimilation of the ring by the central black
hole), to its final Kerr (vacuum) value
$\Omega^{\text{Kerr}}_{\text{H}}(J^{\text{new}}_{\text{H}})=
J^{\text{new}}_{\text{H}}/2{M^{\text{new}}}^2R^{\text{new}}_{\text{H}}$
[that is, {\it after} the adiabatic assimilation of the ring by the
central black hole. Here $J^{\text{new}}_{\text{H}}=J_{\text{R}}$,
$M^{\text{new}}$, and $R^{\text{new}}_{\text{H}}$ are the new
parameters of the resulting Kerr (vacuum) black hole after it
assimilated the orbiting ring]. We use this important observation in
order to generalize the result of Will to the regime of
black-hole-ring configurations in which the central black holes
possess non-zero angular momenta. In particular, it is shown that
the continuity argument (namely, the characteristic {\it smooth}
evolution of the black-hole angular-velocity during an adiabatic
assimilation process of the ring into the central black hole) yields
a concrete prediction for the angular-velocity/angular-momentum
asymptotic functional relation
$\Omega^{\text{BH-ring}}_{\text{H}}=\Omega^{\text{BH-ring}}_{\text{H}}(J_{\text{H}},J_{\text{R}},R\to
R^{+}_{\text{H}})$ of {\it generic} (that is, with
$J_{\text{H}}\neq0$) black-hole-ring configurations. Remarkably, we
find the simple {\it universal} relation
$\Delta\Omega_{\text{H}}\equiv\Omega^{\text{BH-ring}}_{\text{H}}(J_{\text{H}},J_{\text{R}},R\to
R^{+}_{\text{H}})-\Omega^{\text{Kerr}}_{\text{H}}(J_{\text{H}})={{J_{\text{R}}}/{4M^3}}$
for the asymptotic deviation of the black-hole angular-velocity in
the composed black-hole-ring system from the corresponding
angular-velocity of the unperturbed (vacuum) Kerr black hole with
the {\it same} angular-momentum.
\end{abstract}
\bigskip
\maketitle

%]

\section{Introduction}

The gravitational two-body problem has attracted much attention over
the years from both physicists and mathematicians. In particular, it
is highly important to explore the physics of a central black hole
surrounded by an orbiting ring: it is expected that this composed
two-body system may be formed as an intermediate stage in the
gravitational collapse of a compact spinning star to form a black
hole \cite{Ans1,Shap,Shib}. Likewise, the coalescence of two compact
objects may produce a composed black-hole-ring system
\cite{Ans1,Shap,Shib}. In addition to these astrophysical
motivations, it is highly interesting to explore the composed
black-hole-ring system in order to understand how an exterior matter
configuration affects the physical properties of central black holes
\cite{Ans1,Shap,Shib,Will1,Will2}.

The general-relativistic problem of a {\it slowly} spinning black
hole surrounded by a thin orbiting ring was studied perturbatively
by Will \cite{Will1,Will2} (see also \cite{Hod1}). It was shown in
\cite{Will1} that, due to the well-known phenomenon of {\it
dragging} of inertial frames by the orbiting ring, the
angular-velocity $\Omega^{\text{BH-ring}}_{\text{H}}$ \cite{Noteden}
of the central black hole in the composed black-hole-ring system is
{\it no} longer related to the black-hole angular-momentum
$J_{\text{H}}$ by the simple Kerr-like (vacuum) relation
\begin{equation}\label{Eq1}
\Omega^{\text{Kerr}}_{\text{H}}(J_{\text{H}})={{J_{\text{H}}}\over{2M^2R_{\text{H}}}}\
.
\end{equation}
[Here $M$ and $R_{\text{H}}=M+(M^2-J^2_{\text{H}}/M^2)^{1/2}$ are
the mass and horizon-radius of the black hole, respectively]. In
particular, Will \cite{Will1} has demonstrated explicitly that, in
the composed black-hole-ring system, one can have a central black
hole with zero angular-momentum but with a non-zero angular-velocity
\cite{Notecor1}
\begin{equation}\label{Eq2}
\Omega^{\text{BH-ring}}_{\text{H}}(J_{\text{H}}=0,J_{\text{R}},R)={{2J_{\text{R}}}\over{R^3}}\
,
\end{equation}
where $J_{\text{R}}$ and $R$ are respectively the angular-momentum
of the orbiting ring and its proper circumferential radius. To the
best of our knowledge, no exact ({\it analytical}) calculations of
the frame-dragging effect have been performed for {\it generic}
black-hole-ring configurations (that is, for the case of central
black holes with non-negligible angular momenta).

%The lack of detailed analytical results for this important physical
%system may be attributed to the enormous complexity of this
%general-relativistic two-body problem.

\section{The continuous (smooth) behavior of the black-hole angular-velocity}

The main goal of the present paper is to generalize the result
(\ref{Eq2}) of \cite{Will1} to the regime of composed
black-hole-ring configurations in which the central black holes
possess non-zero angular momenta. In particular, we shall use a
simple continuity argument in order to provide a concrete analytical
prediction for the angular-velocity/angular-momentum asymptotic
functional relation
$\Omega^{\text{BH-ring}}_{\text{H}}=\Omega^{\text{BH-ring}}_{\text{H}}(J_{\text{H}},J_{\text{R}},R\to
R^{+}_{\text{H}})$ of generic (that is, with $J_{\text{H}}\neq0$)
central black holes in the composed black-hole-ring system.

Our approach here is based on a {\it continuity} argument for the
behavior of the black-hole angular-velocity in an {\it adiabatic}
process in which the orbiting ring is assimilated (adiabatically
lowered) into the central black hole. In order to demonstrate the
idea, we shall first analyze the analytical relation (\ref{Eq2}) of
\cite{Will1} for the angular-velocity of a zero angular-momentum
($J_{\text{H}}=0$) central black hole.

Let us first consider a sequence of black-hole-ring configurations
with adiabatically varying (decreasing) circumferential radii.
Inspection of Eq. (\ref{Eq2}) reveals that, for a given value of the
ring angular-momentum $J_{\text{R}}$, the central black-hole
angular-velocity {\it increases} as the ring approaches the
black-hole horizon (that is, as $R$ decreases). In particular,
taking the near-horizon limit $R\to R^{+}_{\text{H}}$ in
(\ref{Eq2}), one finds \cite{Notez}
\begin{equation}\label{Eq3}
\Omega^{\text{BH-ring}}_{\text{H}}(J_{\text{H}}=0,J_{\text{R}},R\to
R^{+}_{\text{H}})\to {{J_{\text{R}}}\over{4M^3}}
\end{equation}
for the angular-velocity of the central black hole just {\it before}
it assimilates the ring.

Let us now calculate the new angular-velocity
$\Omega^{\text{Kerr}}_{\text{H}}(J^{\text{new}}_{\text{H}})$ of the
resulting Kerr (vacuum) black hole {\it after} it absorbed the ring.
The adiabatic assimilation of the rotating ring by the central black
hole produces the following changes in the black-hole physical
parameters:
\begin{equation}\label{Eq4}
M\to M^{\text{new}}=M+{\cal E}_{\text{R}}\ \ \ \text{and}\ \ \
J_{\text{H}}=0\to J^{\text{new}}_{\text{H}}=J_{\text{R}}\ ,
\end{equation}
where the energy ${\cal E}_{\text{R}}$ of the rotating ring at the
absorption point $R=R_{\text{H}}$ is given by
\cite{Car,Notezr,Notered,Notecor2}
%,Notezf}
\begin{equation}\label{Eq5}
{\cal E}_{\text{R}}={{J_{\text{H}}}\over{2M^2R_{\text{H}}}}\cdot
J_{\text{R}}\to 0\ \ \ \text{for}\ \ \ J_{\text{H}}=0\  .
%+O(J^2_{\text{R}}/M^3)
\end{equation}
Substituting (\ref{Eq4}) and (\ref{Eq5}) into (\ref{Eq1}), one finds
\cite{Notez}
\begin{equation}\label{Eq6}
\Omega^{\text{Kerr}}_{\text{H}}(J^{\text{new}}_{\text{H}}=J_{\text{R}})={{J_{\text{R}}}\over{4M^3}}\
\end{equation}
for the angular-velocity of the final (vacuum) Kerr black hole
\cite{Notead}.

Comparing the near-horizon asymptotic ($R\to R^{+}_{\text{H}}$)
expression (\ref{Eq3}) for the angular-velocity of the central black
hole in the composed black-hole-ring system just {\it before} the
assimilation of the ring, with the expression (\ref{Eq6}) for the
angular-velocity of the resulting Kerr (vacuum) black hole {\it
after} it assimilated the ring, one concludes that the black hole is
characterized by a {\it smooth} (continuous) evolution of its
angular-velocity during the adiabatic assimilation process. That is,
\begin{equation}\label{Eq7}
\Omega^{\text{Kerr}}_{\text{H}}(J^{\text{new}}_{\text{H}}=J_{\text{R}})=\Omega^{\text{BH-ring}}_{\text{H}}(J_{\text{H}}=0,J_{\text{R}},R\to
R^{+}_{\text{H}})\  .
\end{equation}

\section{The angular-velocity/angular-momentum relation for generic
black-hole-ring configurations}

In the present section we shall analyze the
angular-velocity/angular-momentum asymptotic functional relation
$\Omega^{\text{BH-ring}}_{\text{H}}=\Omega^{\text{BH-ring}}_{\text{H}}(J_{\text{H}},J_{\text{R}},R\to
R^{+}_{\text{H}})$ of generic (that is, with $J_{\text{H}}\neq0$)
central black holes in the composed black-hole-ring system. To that
end, we shall use the characteristic {\it continuity} relation
\cite{Noteconj}
\begin{equation}\label{Eq8}
\Omega^{\text{Kerr}}_{\text{H}}(J^{\text{new}}_{\text{H}}=
J_{\text{H}}+J_{\text{R}})=\Omega^{\text{BH-ring}}_{\text{H}}(J_{\text{H}},J_{\text{R}},R\to
R^{+}_{\text{H}})\
\end{equation}
for the evolution of the black-hole angular-velocity during an {\it
adiabatic} assimilation process of the orbiting ring by the central
black hole.

%To that end, we shall assume that the smooth ({\it continuous})
%evolution of the black-hole angular-velocity [see Eq. (\ref{Eq7})]
%during an {\it adiabatic} assimilation process of the orbiting ring
%into the central black hole is a generic property of the composed
%system. Namely, we shall use the continuity relation \cite{Noteconj}
%\begin{equation}\label{Eq8}
%\Omega^{\text{Kerr}}_{\text{H}}(J^{\text{new}}_{\text{H}}=
%J_{\text{H}}+J_{\text{R}})=\Omega^{\text{BH-ring}}_{\text{H}}(J_{\text{H}},J_{\text{R}},R\to
%R^{+}_{\text{H}})\  .
%\end{equation}

We consider a composed black-hole-ring system which is characterized
by the physical parameters $J_{\text{H}},J_{\text{R}}$, and $R$. The
adiabatic absorption of the ring by the central black hole produces
a final Kerr (vacuum) black hole with the following parameters:
\begin{equation}\label{Eq9}
M\to M^{\text{new}}=M+{\cal E}_{\text{R}}\ \ \ \text{and}\ \ \
J_{\text{H}}\to J^{\text{new}}_{\text{H}}=J_{\text{H}}+J_{\text{R}}\
,
\end{equation}
where the energy ${\cal E}_{\text{R}}$ of the ring at the absorption
point $R=R_{\text{H}}$ is given by
\cite{Car,Notezr,Notered,Notecor2}
\begin{equation}\label{Eq10}
{\cal E}_{\text{R}}={{J_{\text{H}}}\over{2M^2R_{\text{H}}}}\cdot
J_{\text{R}}\ .
\end{equation}
Substituting (\ref{Eq9}) and (\ref{Eq10}) into (\ref{Eq1}), one
finds
\begin{equation}\label{Eq11}
\Omega^{\text{Kerr}}_{\text{H}}(J^{\text{new}}_{\text{H}}=
J_{\text{H}}+J_{\text{R}})={{J_{\text{H}}}\over{2M^2R_{\text{H}}}}+
{{J_{\text{R}}}\over{4M^3}}\
\end{equation}
for the angular-velocity of the final (vacuum) Kerr black hole
\cite{Notead}.

Taking cognizance of Eqs. (\ref{Eq1}) and (\ref{Eq11}), and using
the {\it continuity} argument (\ref{Eq8}) for the evolution of the
black-hole angular-velocity during the {\it adiabatic} assimilation
process of the ring into the central black hole, one finds the
characteristic angular-velocity/angular-momentum asymptotic relation
\begin{equation}\label{Eq12}
%\Delta\Omega\equiv
\Omega^{\text{BH-ring}}_{\text{H}}(J_{\text{H}},J_{\text{R}},R\to
R^{+}_{\text{H}})=\Omega^{\text{Kerr}}_{\text{H}}(J_{\text{H}})+{{J_{\text{R}}}\over{4M^3}}
\end{equation}
for a central black hole of angular-momentum $J_{\text{H}}$ in the
composed black-hole-ring system [Here
$\Omega^{\text{Kerr}}_{\text{H}}(J_{\text{H}})$, as given by
(\ref{Eq1}), is the angular-velocity of a ({\it vacuum}) Kerr black
hole with the {\it same} angular-momentum].

\section{Summary and discussion}

The composed black-hole-ring system is one of the most fundamental
problems in general relativity and astrophysics
\cite{Ans1,Shap,Shib}. This two-body system is characterized by one
of the most intriguing phenomena in general relativity, namely the
{\it dragging} of inertial frames. In a very interesting work, Will
\cite{Will1,Will2} studied this composed system perturbatively in
the regime of {\it slowly} spinning central black holes. It was
shown in \cite{Will1,Will2} that the effect of dragging of inertial
frames by the orbiting ring yields a non-trivial
angular-velocity/angular-momentum relation for the central black
hole. In particular, Will \cite{Will1,Will2} found the non-zero
angular-velocity (\ref{Eq2}) for a central black hole of zero
angular-momentum ($J_{\text{H}}=0$) in the composed black-hole-ring
system.

To the best of our knowledge, in the physical literature there are
no available analytical results for the frame-dragging effect in
{\it generic} black-hole-ring configurations (that is, for central
black holes with non-negligible angular momenta).
%The lack
%of detailed analytical results for this important physical system
%reflects the enormous complexity of this general-relativistic
%two-body problem.
The main goal of the present paper was to generalize the result
(\ref{Eq2}) of Will \cite{Will1,Will2} to the regime of composed
black-hole-ring configurations in which the central black holes
possess non-zero angular momenta.

In particular, we have explored the
angular-velocity/angular-momentum asymptotic functional relation
$\Omega^{\text{BH-ring}}_{\text{H}}=\Omega^{\text{BH-ring}}_{\text{H}}(J_{\text{H}},J_{\text{R}},R\to
R^{+}_{\text{H}})$ of generic black-hole-ring configurations. To
that end, we have used a {\it continuity} argument \cite{Notesm} for
the evolution of the black-hole angular-velocity during a physical
process in which the orbiting ring is adiabatically lowered into the
central black hole. This continuity argument [see Eqs. (\ref{Eq7})
and (\ref{Eq8})] yields the non-trivial (non Kerr-like)
angular-velocity/angular-momentum asymptotic functional relation
(\ref{Eq12}) for {\it generic} (that is, with $J_{\text{H}}\neq0$)
central black holes in the composed black-hole-ring system.

Remarkably, our result (\ref{Eq12}) for the angular velocity of the
perturbed central black hole implies the simple {\it universal}
\cite{Noteuniv} relation
\begin{equation}\label{Eq13}
\Delta\Omega_{\text{H}}(R\to
R^{+}_{\text{H}})={{J_{\text{R}}}\over{4M^3}}\  ,
\end{equation}
where $\Delta\Omega_{\text{H}}(R\to
R^{+}_{\text{H}})\equiv\Omega^{\text{BH-ring}}_{\text{H}}(J_{\text{H}},J_{\text{R}},R\to
R^{+}_{\text{H}})-\Omega^{\text{Kerr}}_{\text{H}}(J_{\text{H}})$ is
the asymptotic deviation of the black-hole angular-velocity
$\Omega^{\text{BH-ring}}_{\text{H}}(J_{\text{H}})$ in the composed
black-hole-ring system from the corresponding angular-velocity
$\Omega^{\text{Kerr}}_{\text{H}}(J_{\text{H}})$ [see Eq.
(\ref{Eq1})] of the unperturbed (vacuum) Kerr black hole with the
same angular-momentum $J_{\text{H}}$. It is worth emphasizing that
the asymptotic relation (\ref{Eq13}) for $\Delta\Omega_{\text{H}}$
is {\it universal} in the sense that it is {\it independent} of the
black-hole angular-momentum $J_{\text{H}}$.

Finally, we would like to end this paper we a {\it conjecture}. In
particular, we would like to suggest a simple (and compact) formula
which generalizes the asymptotic near-horizon result (\ref{Eq13}) to
generic values of the ring radius $R$. To that end, we note that the
simplest \cite{Notenoo} functional relation
$\Delta\Omega_{\text{H}}=\Delta\Omega_{\text{H}}(R)$ which reduces
to (\ref{Eq2}) in the zero angular-momentum $J_{\text{H}}\to0$ limit
\cite{Notetm}, and to (\ref{Eq13}) in the asymptotic near-horizon
$R\to R_{\text{H}}$ limit, is given by \cite{Notedb}
\begin{equation}\label{Eq14}
\Delta\Omega_{\text{H}}(R)={{J_{\text{R}}}\over{4M^3}}\cdot
\Big({{R_{\text{H}}}\over{R}}\Big)^3\ .
\end{equation}
It would be highly interesting to test the validity of our
conjectured relation (\ref{Eq14}) with full non-linear
\cite{Noteemp} numerical computations for generic (that is, with
$J_{\text{H}}\neq0$) black-hole-ring configurations.

\bigskip
\noindent
{\bf ACKNOWLEDGMENTS}
\bigskip

This research is supported by the Carmel Science Foundation. I thank
Yael Oren, Arbel M. Ongo, Ayelet B. Lata, and Alona B. Tea for
stimulating discussions.

%\newpage


\begin{thebibliography}{99}
\bibitem{Noteden} We use the symbol $\Omega^{\text{BH-ring}}_{\text{H}}$ to
denote the angular-velocity of the black-hole horizon in the
composed black-hole-ring system.
\bibitem{Notecor1} This results is valid in the perturbative regime
$J_{\text{R}}/R^2\ll1$. There are also subleading correction terms
of order $O(J^2_{\text{R}}/R^5)$ on the r.h.s of this relation.
\bibitem{Notez} Here we have used the leading-order relation $R_{\text{H}}=2M$ for the
horizon-radius of the zero angular-momentum black hole.
\bibitem{Notered} It is worth emphasizing that the mass-energy of the ring as measured by
asymptotic observers, $\sqrt{g_{00}}\mu$, is completely red-shifted
($\sqrt{g_{00}}=0$) at the absorption point (at the black-hole
horizon).
\bibitem{Notecor2} This results is valid in the perturbative regime
$J_{\text{R}}/R^2\ll1$. There are also small (sub-leading)
correction terms of order $O(J^2_{\text{R}}/R^3,\mu^2/R)$ on the
r.h.s of this relation.
\bibitem{Noteuniv} Note that the relation (\ref{Eq13}) for
$\Delta\Omega_{\text{H}}$ is universal in the sense that it is {\it
independent} of the black-hole angular-momentum $J_{\text{H}}$.
\bibitem{Notenoo} It should be emphasized that the conjectured relation
(\ref{Eq14}) is not unique. For example, the more complicated
expression
$\Delta\Omega_{\text{H}}(R)={{RR_{\text{H}}J_{\text{R}}}\over{M(R^2+J^2_{\text{H}}/M^2)^2}}$
also reduces to (\ref{Eq2}) in the zero angular-momentum
$J_{\text{H}}\to0$ limit (which corresponds to $R_{\text{H}}\to
2M$), and to (\ref{Eq13}) in the asymptotic near-horizon $R\to
R_{\text{H}}$ limit.
\bibitem{Notetm} Note that this limit corresponds to $R_{\text{H}}\to
2M$.
\bibitem{Notedb} Note that $R_{\text{H}}\to 2M$ in the zero angular-momentum
$J_{\text{H}}\to0$ limit, in which case (\ref{Eq14}) reduces to
(\ref{Eq2}). Likewise, $R\to R_{\text{H}}$ in the near-horizon
limit, in which case (\ref{Eq14}) reduces to (\ref{Eq13}).
\bibitem{Noteemp} It is worth emphasizing again that our analysis is
valid in the perturbative regime $J_{\text{R}}/R^2\ll1$.
\end{thebibliography}
\end{document}